\documentclass[aps,prl,reprint,amsmath,amssymb,
superscriptaddress,showpacs,showkeys]{revtex4-1}
\usepackage{graphicx}
\usepackage{dcolumn}
\usepackage{bm}
\usepackage{soul} 
\usepackage{color}
\definecolor{vs}{rgb}{0.1,0.4,0.1}                  

\usepackage{hyperref}
\begin{document}


\title{\protect{\vspace*{5mm}}
Transient effects at resonant light scattering by particles: Anapole as a storage for ``frozen light"?}
%


\author{Sergey E. Svyakhovskiy}
\affiliation{
M. V. Lomonosov Moscow State University, Moscow, 119991, Russia}
\author{Vladimir V. Ternovski}
\affiliation{
M. V. Lomonosov Moscow State University, Moscow, 119991, Russia}
\author{Michael I. Tribelsky}
\email[The corresponding author,]{\mbox{}\\
\mbox{E-mail: mitribelsky\_at\_gmail.com} (replace ``\_at\_" by @).}
\affiliation{
M. V. Lomonosov Moscow State University, Moscow, 119991, Russia}
\affiliation{National Research Nuclear University MEPhI (Moscow Engineering Physics Institute), Moscow, 115409, Russia}
\affiliation{Landau Institute for Theoretical Physics RAS, Chernogolovka, Moscow Region 142432, Russia}
\affiliation{RITS Yamaguchi University, Yamaguchi, 753-8511, Japan}
\begin{minipage}[T]{\textwidth}
\flushright{
\textit{Is there anybody going to listen to my story\\
All about the} wave that \textit{came to stay}?\\
Based on the Beatles
}
\end{minipage}
\date{\today}

\begin{abstract}
Transient effects at excitation of high-Q resonant modes at a leading front of a laser pulse and their decomposition at its trailing edge during the pulse scattering by a particle are discussed. 
The main attention is paid to the nonradiating (anapole) mode excited in a high-index cylinder. The problem is studied both analytically and numerically. It is shown that the anapole is a steady but essentially nonequilibrium mode, which can exist if and only if the host particle is irradiated by a continuous incident wave. As soon as the incident wave is switched off the anapole becomes a radiating mode and collapses very fast owing to extensive radiative losses. It ruins the hopes to employ the anapole-like excitations as storages for the ``frozen light."
\end{abstract}


\maketitle
Systematic study of light scattering by small particles, originated more than hundred years ago in classical works by such giants as Clebsch, Lamb, Lorenz,  Rayleigh, Debye and others (see the excellent historical review of the problem in Kerker's book~\cite{Kerker2013}), still remains one of the most appealing topics of electrodynamics. Moreover, nowadays the interest in this phenomenon is growing up incrementally owing to its numerous applications in various specific fields, ranging from medicine and biology to telecommunications, data processing and storage, etc.~\cite{Novotny:Book:2006,Klimov:Book:2014}. It stimulates the extensive study of the issue, revealing new, and often unexpected, peculiarities in the old and seemingly well-known problem. Among them the anomalous scattering by metallic particles with low dissipative losses~\cite{TribelskyJETP1984:ResonantScat,tribelskii2006anomalous2780463} and its analog at the light scattering by high-index dielectric particles~\cite{Kuznetsov_Optically_2016,Tribelsky_Mirosh_2016} should be mentioned. A key feature of this type of the scattering is a complicated topological structure of the vector Poynting field in the near field wave zone and its sharp dependence on small variations of the incident wave frequency~\cite{tribelskii2006anomalous2780463,lukyanchuk2014light6656959}.

However, the higher $Q$-factor, the longer transient to the steady-state. Thus, the discussed resonances should be characterized by transient processes lying within the range of experimentally-resolvable time-scales. It is natural to expect that these transient processes should be accompanied by qualitative changes of the topological structure of the electromagnetic and vector Poynting fields. Nonetheless, despite fast processes are the subject attracting close attention of researches all around the world~\cite{ciappina2017attosecond}, much to our surprise, we are not aware of a single publication devoted to the problem indicated above. In the present Letter we are trying to begin filling the gap. The specific problem, we are focused on, is the transient processes related to the so called \emph{anapole} excitations.

The anapole is a nonradiating resonant mode excited in a scattering particle by a plane incident electromagnetic wave. The total suppression of the scattering for this mode occurs owing to the destructive Fano resonance~\cite{Fano:PR:1961}. To the best of our knowledge the anapole was introduced in paper ~\cite{Miroshnichenko_Nonradiating_2015}. To unveil the physics of the phenomenon the authors of Ref.~\cite{Miroshnichenko_Nonradiating_2015} have developed an alternative description of the nonradiating nature of the anapole, presenting the excitation as a superposition of the Cartesian electric and toroidal modes, excited in the scattering particle, whose radiation cancel each other.

Immediately after that publication the anapole and anapole-like excitations have attracted lots of attention of the best scientific teams in different countries~\cite{Papasimakis_Electromagnetic_2016}. Among other reasons such a boom is associated with the expectations that the anapole could be a storage for the ``frozen light" i.e., localized optical-frequency electromagnetic oscillations~\cite{Wiersma_Localization_1997}.

Indeed, since for the anapole the radiative losses are totally suppressed, it seems that in the non-dissipative limit the anapole could exist forever, even if the exited it incident wave is switched off. In other words, the excitation could be a storage for the electromagnetic oscillations in the same manner as a capacitor is a storage for electrical charges. If this is the case, it opens a door to the creation of entirely new optical memory cells and other revolutionary technologies.

Nonetheless, despite the apparent importance of the problem, the experimental observations of the anapole has been done just in the cases, when the exciting it laser pulse is not over~\cite{Miroshnichenko_Nonradiating_2015,Papasimakis_Electromagnetic_2016}. Evidence of a ``stand alone" anapole, existing for a reasonable time after the exiting laser pulse is over, has not been reported yet.

{Note also a paradox related to the nonradiating nature of the anapole. If a scatterer in a nonradiating mode does not emit any radiation, then, thanks to the invariance of the Maxwell equations in the nondissipative media against the reversion of time, it cannot receive any radiation from outside too. It means that the anapole cannot be excited, at least as long as a plane continuous electromagnetic wave is a concern. These arguments are in deep contradiction with the solid experimental evidence of such an excitation in the case of weakly dissipating silicon disk~\cite{Miroshnichenko_Nonradiating_2015}.}

In the present Letter we are going to reveal the physical grounds of the failure in quest of the stand alone anapole and resolve the paradox. The point is that t\emph{he stand along anapole merely does not exist} --- as soon as the exiting anapole incident radiation is off the anapole becomes a radiating excitation and quickly dies out owing to extensive radiative losses.

How does this behavior agree with the nonradiating nature of the anapole? To find the answer we have to recover the physical grounds for the destructive Fano resonances. Following the reasoning of Fano himself~\cite{Fano:PR:1961}, the resonances are explained as a result of interference of two scattered waves (\emph{partitions}): background and resonant~\cite{Miroshnichenko:RMP:2010}. For the problem in question the background partition corresponds to the scattering by the same particle but made of the hypothetical perfect electric conductor (PEC), while the resonant partition is a resonant Mie mode~\cite{Tribelsky_Mirosh_2016}. The same result follows from the alternative description, developed in Ref.~\cite{Miroshnichenko_Nonradiating_2015}, where the cancelation of the radiation occurs due to the interference of the two modes --- Cartesian electric and toroidal.

{This picture is valid at the steady state, corresponding to the scattering of a continuous wave. What happens if the incident wave is switched on/off abruptly? A qualitative answer to this question follows from simple dimensional analysis. By definition, the background scattering does not have any singled out frequencies. It means that the only quantity with the dimension of time, related to this scattering is the the period of the field oscillations in the incident wave: $2\pi/\omega$. This is the shortest characteristic time-scale in the problem and just this time-scale determines the characteristic relaxation time for the background partition at abrupt changes of the amplitude of the incident wave.}

{Regarding the story about the resonant partition, it is completely different. This partition has the resonant frequency and the \emph{linewidth} $\Gamma$. Then, the characteristic relaxation time for the resonant mode is of the order $1/\Gamma \gg 2\pi/\omega$ (we are dealing with high-$Q$ resonances). }

{The difference in the transient processes for the two partitions is related to their different physical origin. The background partition corresponds to the solution of the axillary problem of the scattering by a PEC cylinder. The PEC does not allow any electromagnetic field to get in. Then, the scattering is produced entirely owing to the surface currents induced by the incident wave.

On the other hand, in the actual problem, thanks to the boundary conditions, any abrupt change of the incident wave should produce the corresponding immediate change in the tangential components of $\mathbf{E}$ and $\mathbf{H}$ in a thin boundary layer, adjacent to the surface of the cylinder. It generates a substantial surface displacement current, changing dramatically the current responsible for the PEC cylinder scattering. In contrast, the field in the bulk of the cylinder initially is weakly affected --- it takes a time for the surface perturbations to penetrate into the bulk.}

{Thus, at abrupt switching of the incident wave on/off the background partition follows these variations \emph{practically immediately}, while the resonant partition can do that only with a considerable delay. It destroys the balance between the background and resonant partitions required for the destructive interference. The partitions cannot cancel each other any more and the excitation becomes \emph{radiating}. This reasoning explains why the anapole may be excited by the leading front of an incident pulse and gives grounds to expect its collapse due to radiative losses at the trailing edge of the pulse.}
%

Let us elaborate these arguments, based on the case elucidated in Ref.~\cite{Miroshnichenko_Nonradiating_2015}. To simplify the analysis we consider the 2D problem of the scattering by the cylinder whose axis is perpendicular to the plane of oscillations of vector $\mathbf{E}$ of the incident wave and to the wave vector of this wave (TE polarization, normal incidence), however the results obtained are generic and valid for other geometries (including the 3D sphere scattering) too. The steady version of the problem is exactly solvable~\cite{Kerker2013}. According to the solution, the scattered field and the field within the cylinder both are presented as infinite series of partial multipolar waves of the $\ell$th order ($-\infty \leq \ell \leq \infty$). The corresponding partial fields read as follows:
\begin{eqnarray}
  & & \frac{\mathbf{E}_\ell^{(TE)}}{E_0} = (-i)^{\ell+1}e^{i \ell \phi }d_\ell \left\{i \ell \frac{ J_\ell(m\rho)}{m\rho},-J_\ell'(m\rho),0\right\},\label{eq:Ein_TE} \\
  & & \frac{\mathbf{H}_\ell^{(TE)}}{H_0}=-m(-i)^\ell e^{i \ell \phi } d_\ell \left\{0,0,J_\ell(m\rho) \right\}, \label{eq:Hin_TE}
\end{eqnarray}
within the cylinder, and
\begin{eqnarray}
  & &\!\!\!\!\!\! \frac{\mathbf{E}_\ell^{(TE, s)}}{E_0} = -(-i)^{\ell+1}e^{i \ell \phi }a_\ell \left\{i \ell \frac{ H^{(1)}_\ell(\rho)}{\rho},-H^{(1)'}_\ell\!\!(\rho),0\!\right\}\! ,\label{eq:Eout_TE} \\
  & & \frac{\mathbf{H}_\ell^{(TE, s)}}{H_0}= (-i)^\ell e^{i \ell \phi } a_\ell \left\{0,0, H^{(1)}_\ell(\rho) \right\}. \label{eq:Hout_TE}
\end{eqnarray}
outside it. Here $E_0$ and $H_0$ are, respectively, the amplitudes of the electric and magnetic fields in the incident wave, whose wave vector $\mathbf{k}$ is aligned {antiparallel} to $x$-axis; $\{X_r,X_\phi,X_z\}$ denote the components of any vector $\mathbf{X}$ in the cylindrical coordinate frame with the \mbox{$z$-axis} directed along the axis of the cylinder; $\rho \equiv rk$; $k = \omega/c$ is the wavenumber of the incident wave in a vacuum; $m \equiv \sqrt{\varepsilon}$ is the refractive index of the cylinder (in what follows, $m$ is supposed to be a purely real quantity --- the nondissipative limit); $\varepsilon$ is its permittivity; $J_\ell(z)$ and $ H^{(1)}_\ell(z)$ stand for the Bessel and Hankel functions of the first kind, respectively, and the prime denotes the derivative with respect to the entire argument of a function. The cylinder is regarded nonmagnetic, so that its permeability $\mu$ equals unity.

%

Thus, all individual information about a specific case of the wave scattering is hidden in the values of the refractive index $m$ and the scattering coefficients $a_\ell$, $d_\ell$. The latter are given by the expressions: 
\begin{eqnarray}
 & & a_\ell = \frac{F_\ell}{F_\ell + iG_\ell},\;\; d_\ell =  \frac{i}{F_\ell + iG_\ell}, \label{eq:ad}\\
 & & F_\ell  = \frac{\pi q}{2}\Big(mJ_\ell(mq)J_\ell'(q) - J_\ell'(mq)J_\ell(q)\Big), \label{eq:F} \\
 & & G_\ell = \frac{\pi q}{2}\Big(mJ_\ell(mq)N_\ell'(q) - J_\ell'(mq)N_\ell(q)\Big), \label{eq:G}
\end{eqnarray}
where $N_\ell(z)$ are the Neumann functions; $q = kR$ stands for the \emph{size parameter} and $R$ is the radius of the cylinder. Note, that at purely real $m$ the quantities $F$ and $G$ are also purely real. Note also, that $|d_\ell| = 1$ at $m=1$ thanks to identity $J_\ell(z)N_\ell'(z) - J_\ell'(z)N_\ell(z) \equiv 2/(\pi z)$. Then,
the departure of $|d_\ell|$ from unity could be a quantitative measure of the rate of the $\mathbf{E}$ field enhancement (suppression) within the particle. The same role plays the departure of $m|d_\ell|$ for field $\mathbf{H}$, see Eqs.~\eqref{eq:Ein_TE}, \eqref{eq:Hin_TE}.

The destructive Fano resonances correspond to the roots of the equation $F_\ell(m,q) = 0$, when $a_\ell$ vanishes. Since, as it is straightforwardly seen from Eqs.~\eqref{eq:ad}--\eqref{eq:G}, coefficients $d_\ell$ never vanish, it means that at the points of the destructive resonances the resonant multipole does not contribute anything to the scattering field, but still has a finite field amplitude within the cylinder, i.e., becomes nonradiating.

To move further we have to be more specific. To this end let us inspect the case $\ell = \pm 1$ and $m=4$, which corresponds to the anapole introduced in Ref.~\cite{Miroshnichenko_Nonradiating_2015}. The first root of the equation $F_1(4q)=0$ is $q \approx 1.0420$.
%
%
%
%

To inspect the dynamical behavior of the scattering problem we employ a numerical integration of the complete set of Maxwell's equations with the standard boundary conditions at the surface of the cylinder~\cite{Kerker2013}. To this end our own code has been developed. Its details will be discussed elsewhere. Here we just would like to stress that the code has been carefully tested against the asymptotical convergence of the generated time-dependent field patterns to the exact analytical solution at a broad domain of variations of the problem parameters and exhibited the excellent agreement with the exact solution.

The formulation of the corresponding physical problem is as follows. We suppose that the pulse duration $\tau$ is much larger than the characteristic microscopic times, so that the simple connection between the instant values of the the electric field $\mathbf{E}(t)$ and the induction $\mathbf{D}(t)$ of the form  $\mathbf{D} = \varepsilon(\omega) \mathbf{E}$ still remains valid. 
We also suppose, that $\omega \tau \gg 1$, i.e., the incident wave may be regarded as monochromatic. In what follows the range of $\omega$-variations is extremely narrow. It makes possible to neglect the dispersion of the permittivity, supposing $\varepsilon = const$. As it has been mentioned above, to make our problem analogous to that presented in Ref.~\cite{Miroshnichenko_Nonradiating_2015} we set $\varepsilon =  16$ ($m = 4$).

%

We employ a rectangular laser pulse, so that the amplitude of the incident wave is zero at $t < 0$, a constant at $0 \leq t \leq \tau$ and zero at $t > \tau$. We suppose that $\tau \Gamma \gg 1$, where $\Gamma$ is the linewidth of the dipole resonant mode (see below). It allows the transient processes to the steady field pattern to be completed during the laser pulse. In this case the field dynamics at the leading edge of the pulse makes it possible to study the transient to the steady state, when the incident field is abruptly switched on. The one at the trailing edge exhibits the opposite process, when the incident wave is abruptly switched off. Finally, it is convenient to introduce the dimensionless time $\theta = \omega_0 t$ and frequency $\Omega = \omega/\omega_0$, where $\omega_0 $ is the eigenfrequency of the resonant dipole mode corresponding to the maximum of its resonance line ($q_{\rm max} \approx 0.9153$) and to normalize the spatial scale over $R$.

%

\begin{figure}
  \centering
  \includegraphics[width=0.5\textwidth]{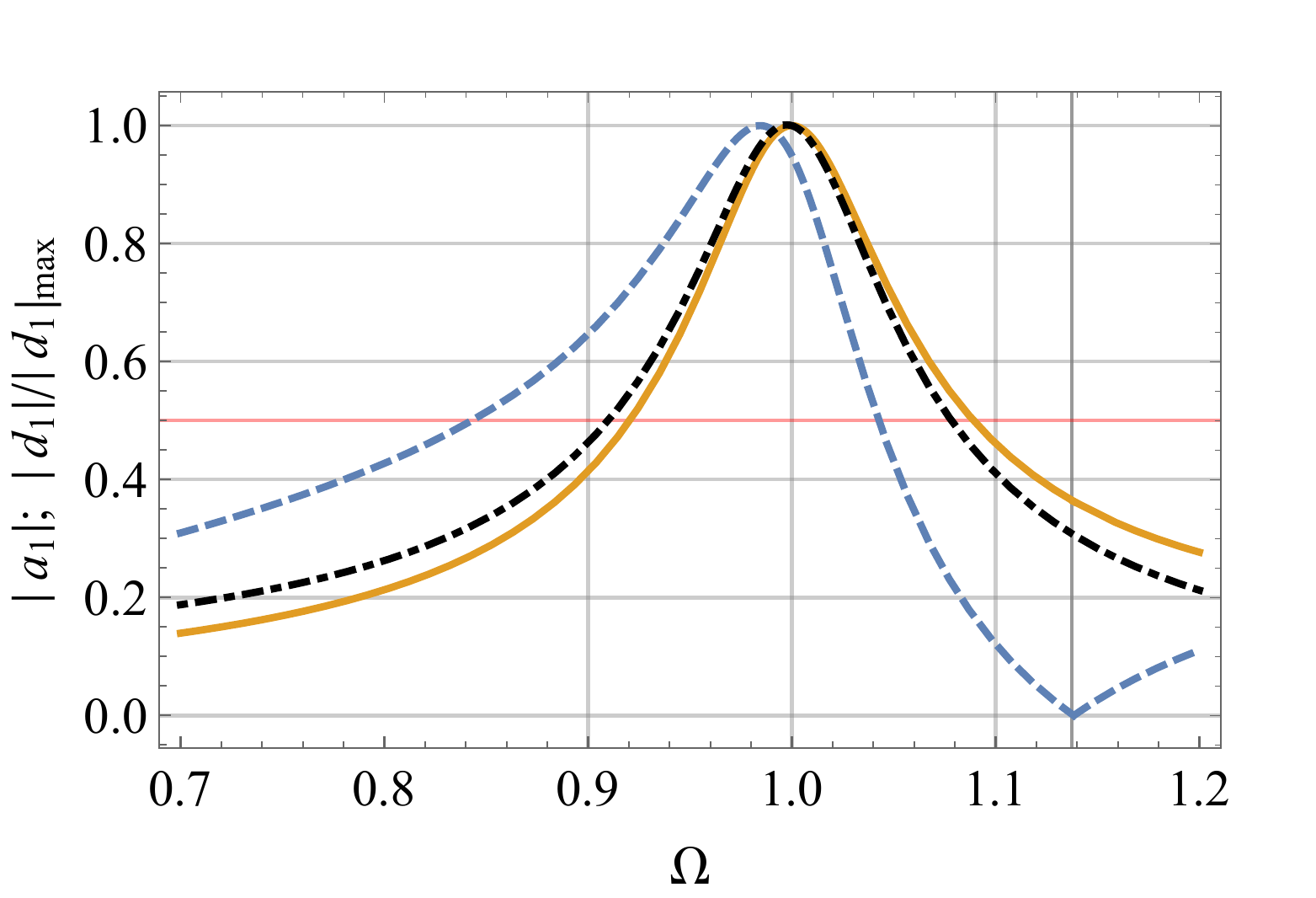}
  \caption{The resonance lines $|a_1|$ and $|d_1|$ for the exact solution of the diffraction problem (solid brown and dashed blue lines, respectively) and the fitting of $|d_1(\Omega)|$ with the line for a pendulum (black dot-dashed line). Both the profiles for $|d_1|$ (the actual and the fitted) are normalized over the corresponding maximal values; $\Omega = 1.13843...$ (marked with a vertical line) is the resonant anapole frequency.}\label{fig:Pendulum-Anapole_Lines}
\end{figure}

The specific frequency of the incident wave is selected so that $a_1(\Omega)=0$ (the anapole resonant frequency). For the given value of $m=4$ it gives rise to $\Omega = 1.13843...\; (q=1.04199...)$, see Fig.~\ref{fig:Pendulum-Anapole_Lines}.

\begin{figure}
  \centering
  \includegraphics[width=.5\textwidth]{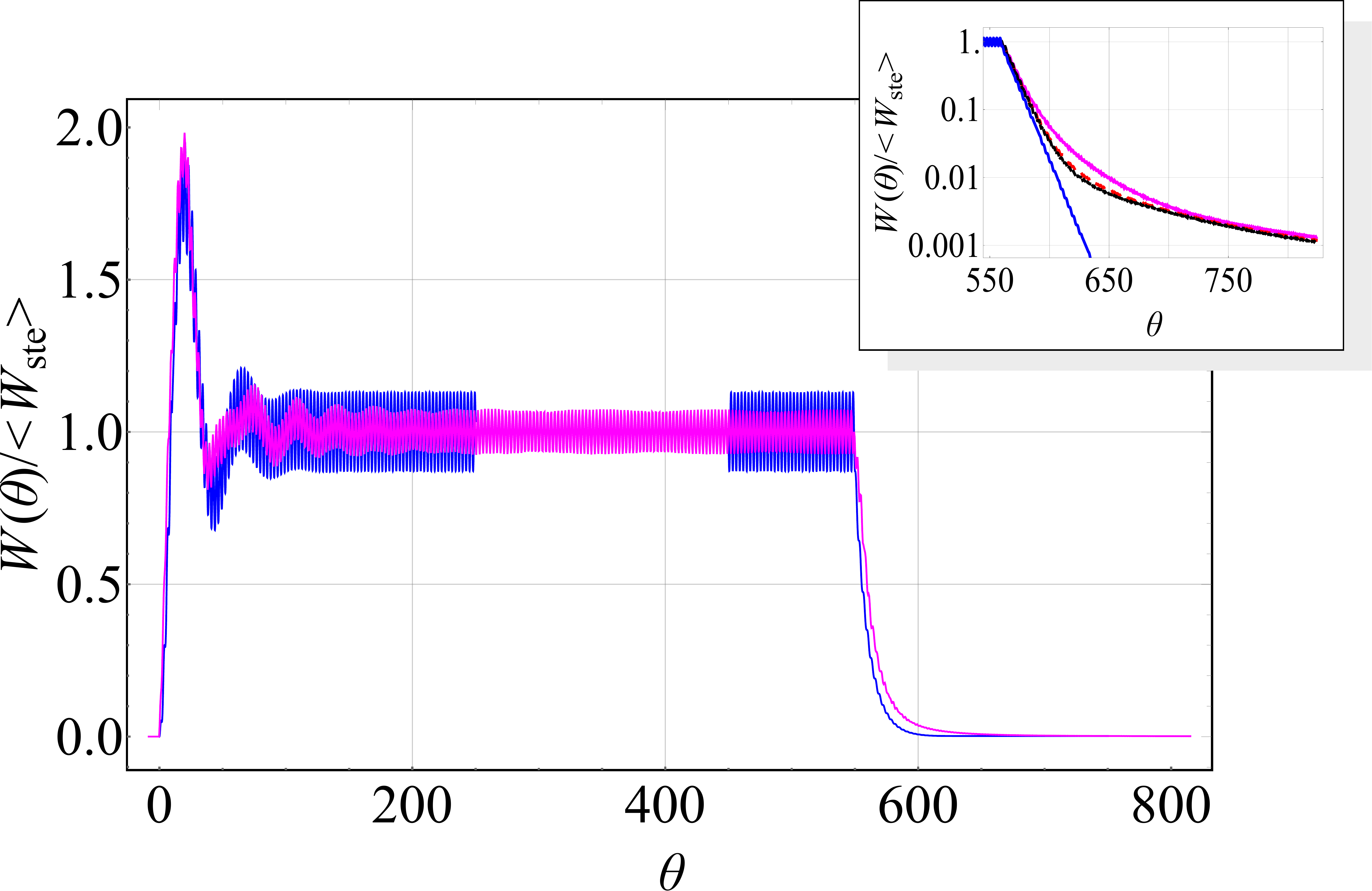}
  \caption{The normalized instant net energy in the irradiated cylinder (magenta) and a simple pendulum (blue). The inset shows the tails of the decay processes, where two more curves for the actual problem with $\Omega =  1.06922...$ (dashed orange) and $\Omega=1$ (dotted black) are added. See the text for details.}\label{fig:Pendulum-Anapole_W}
\end{figure}

To characterize the transient processes quantitatively we have calculated the instant value of the net electromagnetic energy stored in the cylinder, $W(\theta)$:
\begin{equation}\label{eq:W}
  W(\theta)=\frac{1}{8\pi}\int_{0}^{R}rdr\int_{0}^{2\pi}d\varphi\left(\varepsilon \mathbf{E}^2 + \mathbf{H}^2\right),
\end{equation}
where $\mathbf{E}(\theta)$ and $\mathbf{H}(\theta)$ are \emph{real} quantities. The plot $W(\theta)$ normalized over $\left\langle W_{\rm ste} \right\rangle $ is shown in Fig.~\ref{fig:Pendulum-Anapole_W}. Here  $\left\langle W_{\rm ste} \right\rangle$ designates $W(\theta)$ averaged over the period of the incident wave oscillations for the steady scattering process. Time $\theta$ begins to be counted from the moment when the leading front of the incident pulse hits the surface of the cylinder. In the introduced dimensionless units the duration of the pulse $\tau\omega_0= 550$. It is seen clearly that a collapse of the anapole begins immediately after the incident pulse is over.

%

To understand the main features of the numerically obtained dependence $W(\theta)$ we introduce a simple toy model --- forced oscillations of a linear pendulum:
\begin{equation}\label{eq:d_pendulum}
  d_{\theta \theta} + \Gamma d_\theta + d = A \exp(i\Omega\theta).
\end{equation}
To describe the excitation of the anapole by the incident wave Eq.~\eqref{eq:d_pendulum} should be supplemented with the initial conditions $d(0)=d_\theta(0)=0$. To describe the collapse of the anapole after the incident radiation is switched off the initial conditions correspond to that for the steady forced oscillations of the pendulum, while the right hand side of Eq.~\eqref{eq:d_pendulum} should be set to zero. Eq.~\eqref{eq:d_pendulum} is exactly integrable. The integration is trivial, but the expressions describing the dynamics of the \emph{switching on} and \emph{switching off} processes are cumbersome and will not be presented here.

{To adjust the parameters of the toy and actual problems for their quantitative comparison we fit the shape of the resonance line for the pendulum to the one for $|d_\ell|$ in the actual problem, so that the linewidths for the two lines become equal. At the selected values of the problem parameters it gives rise to $\Gamma \approx 0.0967$. The fitting is shown in Fig.~\ref{fig:Pendulum-Anapole_Lines} (it is worth adding that the accuracy of the fittingt increases with an increase in the value of the refractive index). 
Both profiles $|d_1(\Omega)|$ (actual and fitted) are normalized over their maximal values. For the actual diffraction problem $|d_1|_{\rm max} \approx 2.449$. Thus, the actual values of $|d_1|$ and $m|d_1|$ at the anapole resonance point are 0.88884... and 3.55536..., respectively. The next largest coefficient in the multipolar expansion at the given value of $\Omega$ is $d_2$, whose modulus equals $|d_2| \approx 0.250656 \ll |d_1|$. In other words, the anapole mode still makes the leading contribution to the field within the cylinder and while the characteristic value of $\mathbf{E}$ for the anapole is of the order of $E_0$, for $\mathbf{H}$ this quantity is larger then that for the incident wave.}
%

The total energy of the pendulum $W$ is $(1/2)[({\rm Re}\, d_\theta)^2 + ({\rm Re}\, d)^2]$. The dependence $W(\theta)/ \left\langle W_{\rm ste} \right\rangle$, superimposed on the corresponding dependence for the actual problem, is shown in Fig.~\ref{fig:Pendulum-Anapole_W}. Here in the agreement with the actual problem $\left\langle W_{\rm ste} \right\rangle$ designates the average of $W(\theta)$ over the period of the asymptotical steady-state forced pendulum oscillations. At the selected value of $\Gamma$ at $\theta = 250$ the pendulum oscillations become practically steady and the corresponding plot is cut off. Then, the dynamics of the decay of the oscillations with a part of the steady-state, adjacent to the switching off moment is shown.

It should be stressed that the linewidth $\Gamma$ is the only fitting parameter of the toy model. Nonetheless, the agreement of the dynamics of $W(\theta)$ in the actual problem and in the toy model is quite impressive \footnote{The smaller amplitude of the high-frequency oscillations of $W(\theta)$ in the actual problem relative to those in the toy model is an artefact related to the rough sampling of the huge database of the numerics in order to present it graphically --- owing to the sampling the actual extrema of the rapid oscillations cannot be resolved. The cross check against the exact steady analytical solution, when the oscillations can be calculated explicitly shows that their amplitude in the actual problem and in the steady state of the pendulum equal each other.}. A little difference is observed just in the decay rate. It is hardly noticeable in the linear scale, but more pronounced in the logarithmic one, see the inset in Fig.~\ref{fig:Pendulum-Anapole_W}. While the toy model exhibits the pure exponential decay, 
in the actual problem the decay gradually departures from the exponential law, becoming less rapid. However, this slower decay is not a specific feature of the anapole mode. To check that we have simulated the decay dynamics in the actual problem for $\Omega = 1$ (the maximum of the resonance line) and for $\Omega = 1.06922...\;(q = 0.978633)$, lying just in between $\Omega = 1$ and the resonant anapole frequency. The three curves demonstrate a similar behavior, approaching asymptotically the same law (though the decay of the anapole mode is a bit slower). In our view this subtle difference in the decay dynamics in the actual and toy problems hardly can have any practical importance since it becomes pronounced, when more than 90\% of the energy stored in the cylinder is already irradiated out.

%
%
The proposed toy model makes it possible to explain the key features of the observed dynamics of the actual problem in a very simple manner and get a deeper insight of it. Specifically, when the drive is abruptly switched on, in addition to the forced oscillations with frequency $\Omega$ the free oscillations with the eigenfrequency $\Omega=1$ are excited. Since there is a mismatch between the two frequencies, see Fig.~\ref{fig:Pendulum-Anapole_Lines}, it results in beatings in the net energy with the frequency approximately equal to $|\Omega-1|$. The beatings decay with the characteristic time-scale $\sim 1/\Gamma$. It gives rise to an oscillatory relaxation of the average (over the period of the fast underling oscillations) energy stored in the cylinder $\left\langle W(\theta)\right\rangle$, so that initially the the energy ``overfills" the cylinder and reaches a pronounced peak, whose value is considerably larger than $\left\langle W_{\rm ste}\right\rangle$. Then, the excess of the energy is irradiated from the cylinder, $\left\langle W(\theta)\right\rangle$ becomes lower than $\left\langle W_{\rm ste}\right\rangle$, and so on. The simulation with other values of $\Omega$ shows that when $\Omega$ approaches unity the period of the beating increases. Since their damping rate, determined by $\Gamma$, remains fixed, it results in suppression of the amplitude of the beating. Eventually, at $\Omega = 1$ the beatings disappear and $\left\langle W(\theta)\right\rangle$ becomes a smooth, monotonic function.
%


In contrast, at the trailing edge of the pulse, when the gain is switched off instantly, just the free oscillations with $\Omega=1$ are excited. Thus, the maximal power irradiated by the cylinder corresponds to the very beginning of the decay process, then it gradually decreases in the course of time, so that $\left\langle W(\theta)\right\rangle$ is a monotonically decreasing function at any frequency of the switched off drive.
%

It is worth presenting some numerical estimates. As it is seen from Fig.~\ref{fig:Pendulum-Anapole_W}, both transient processes (at the leading and trailing edges of the pulse) last about 50 dimensionless units. It lies within the experimentally resolved time-scales even if the resonant anapole frequency corresponds to the visible range of the spectrum, to say nothing about far IR and radio diapasons.


Note, that if nonlinear effects with the threshold lying above $\left\langle W_{\rm ste}\right\rangle$ but smaller than the peak value of $\left\langle W(\theta)\right\rangle$ are incorporated into the problem, then the pronounced peak of the field intensity in the cylinder at the leading front of the incident pulse may be effectively employed to design new nanodevices generating ultrashort laser pulses.

In conclusions we may say the following. (i) Unfortunately, the anapole and anapole-like excitations cannot serve as storages for the frozen light. They are essentially nonequilibrium and may exist if and only if the host particle is irradiated by a continuous wave. They die off rapidly owing to radiative losses as soon as the incident wave is switched off. (ii) The main features of the dynamics of excitation and collapse of the anapole may be explained within the framework of the toy model of forced oscillations of a simple pendulum.

In a more broad context we should stress that transient processes at resonant Mie's scattering are accompanied with dramatic changes in the topological structure of the electromagnetic field within the scattering particle and its near field zone, whose detailed discussion will be presented elsewhere. In addition to purely academic interest it could find plenty of applications in technologies, such as telecommunications, data storage and processing, optical computers, etc. We believe our results may stimulate further study in this appealing field.


MT acknowledges the financial support of Russian Foundation for Basic Research (Grant 17-02-00401) in part of the analytical modeling and Russian Science Foundation (Grant 14-19-01599) in part of the computer simulation.

\bibliography{Anapole}
\end{document}